\documentclass[twocolumn,reprint,nofootinbib,showpacs]{revtex4}
\usepackage[T1]{fontenc}
\usepackage[english]{babel}
\usepackage[latin9]{inputenc}
\usepackage{verbatim}
\usepackage{amstext}
\usepackage{color}
\usepackage{graphicx,psfrag}
 \usepackage{amssymb}

\def\eg{\emph{e.g.~}}
\def\ie{\emph{i.e.~}}

\def\beq{\begin{equation}}
\def\eeq{\end{equation}}
\def\beqa{\begin{eqnarray}}
\def\eeqa{\end{eqnarray}}
\def\beg{\begin{lyxgreyedout}}
\def\eeg{\end{lyxgreyedout}}

\begin{document}

%

\title{Cyclic spacetime dimensions}
\author{Donatello Dolce}
\email{donatello.dolce@unicam.it}
\affiliation{University of Camerino, Piazza Cavour 19F, 62032 Camerino, Italy.}
\date{\today}

\begin{abstract}
  Every system in physics is  described in terms of interacting elementary particles characterized by modulated spacetime recurrences. These intrinsic periodicities, implicit in undulatory mechanics, imply that every free particle is a reference clock linking time to the particle's mass, and every system is formalizable by means of modulated elementary spacetime cycles. We propose a novel consistent relativistic formalism based on intrinsically cyclic spacetime dimensions, encoding the quantum recurrences of elementary particles into spacetime geometrodynamics.   The advantage of the resulting theory is a formal derivation of quantum behaviors from relativistic mechanics, in which the constraint of intrinsic periodicity turns out to quantize the elementary particles; as well as a geometrodynamical description of gauge interaction which, similarly to gravity, turns out to be represented by relativistic modulations of the internal clocks of the elementary particles. The  characteristic classical to quantum correspondence of the theory brings novel conceptual and formal elements to address fundamental open questions of modern physics.  
\end{abstract}

\pacs{11.90.+t,03.70.+k,03.65.-w}

\maketitle

\paragraph*{\bf Introduction}---
According to the Standard Model, natural phenomena are described in terms elementary particles and their relativistic interactions (including gravity). In addition to this, in agreement with our empirical observations of nature, undulatory mechanics tell us that to every particle of local four-momentum $\bar k_{\mu}=\{\bar \omega, -\mathbf{\bar{k}}\}$ corresponds an instantaneous spacetime recurrence $T^{\mu}=\{T, \vec \lambda\}$ of fundamental  topology $\mathbb S^{1}$. This is, in fact, the Lorentz projection $\Lambda \rightarrow \gamma T - \gamma \vec \beta \cdot \vec \lambda$ of the Compton proper-time periodicity $\Lambda = 2 \pi / \bar m $ of a rest particle of mass $\bar m$ --- in natural units ($\hbar = c = 1$).  In modern physics the recurrence of elementary particles are indeed represented as ``periodic phenomena'' \cite{Broglie:1924,1996FoPhL}, such as  waves or phasors (wave-particle duality), in which spacetime enters as angular variables: their periodicities are locally fixed through the Planck constant by the kinematical states, according to the phase harmony condition $\bar m \Lambda  \equiv 2 \pi  \rightarrow  {\bar{k}}_{\mu}  T^{\mu}\equiv 2 \pi $, where $\bar \omega = \gamma \bar m$ and $\bar k_{i} = \gamma \beta_{i}  \bar m$.  \emph{Thus, every system in nature is formalizable in terms of elementary spacetime cycles and their modulations} \cite{Dolce:2009ce,Dolce:tune,Dolce:AdSCFT}.  
 As also noted for instance by R. Penrose, under such an assumption  of Intrinsic Periodicity (IP) \emph{every isolated particle is a relativistic reference clock}  \cite{Dolce:2009ce,Penrose:cycles}, \ie as ``clocks directly linking time to a particle's mass'' \cite{Lan01022013}.  In fact, ``{a  relativistic clock [is] a phenomenon passing periodically through identical phases}'' \cite{Einstein:1910}. Actually, these 
 ``ticks''  of the internal clocks of the elementary particles (Compton clocks) can be indirectly observable experimentally \cite{2008FoPh...38..659C,Lan01022013}, and in principle used to define our time axis similarly to the atomic clocks. This suggests a relational interpretation of the flow of time. An isolated system (universe) composed by a single particle has a perfectly homogeneous cyclic evolution (persistent ``ticks''), as a pendulum in the vacuum. Nevertheless, a  system composed by more particles is ergodic as the elementary periodicities are in general not rational each others; if we also consider interactions and the consequent modulations of periodicities, we find that the evolution of this non elementary system of elementary cycles is chaotic, according to our physical observations. Every instant in time is therefore characterized by a unique combination of the phases of the elementary cycles associated to the particles constituting the system, similarly to the reference temporal cycles of a stopwatch or of a calendar. The infinite spacetime periodicities of very low energy massless particles (\eg photons and gravitons have infinite Compton worldline periodicities) set the underlying non-compact reference spacetime framework for the ordinary causal relational description of retarded events in relativity. 

The simple question, never considered before, that we want to address is: \emph{what happens if we impose to a particle its natural local modulated spacetime recurrence as a constraint}?\footnote{This means to promote IP to a general principle of physics.} Remarkably, the result is a consistent unified description of fundamental aspects of modern physics based on cyclic spacetime dimensions. Similarly to Bohr's atom or a 'particle in a box', such a constraint of IP is in fact a quantization condition. We will see that the resulting cyclic dynamics formally match ordinary relativistic Quantum Mechanics (QM) in both the canonical and Feynman formulations \cite{Dolce:2009ce,Dolce:tune,Dolce:AdSCFT}. The local retarded variations of four-momentum of avery particle interacting in a given interaction scheme can be equivalently described, through the Planck constant, in terms of corresponding local retarded modulations of elementary spacetime cycles. In this way we will see that gauge interactions can be derived from spacetime geometrodynamics similarly to gravity and Weyl's original proposal \cite{Weyl:1918ib,Blagojevic:2002du}. The assumption of IP is also an interesting, novel interpretational key of the fundamental properties of modern integrable theories. 

\paragraph*{\bf Cyclic spacetime geometrodynamics}--- We represent mathematically the elementary spacetime cycles  of a bosonic particle (``periodic phenomenon'')  in terms of compact spacetime dimensions of length  $T^{\mu}$ with Periodic Boundary Conditions (PBCs).  In this way, a free elementary bosonic system of fundamental mass $\bar m$, in the rest or in a generic reference frame, is represented, respectively, by the equivalent (upon Lorentz transformations) actions 
\begin{equation}
  {\mathcal{S}}^{\Lambda} = \int^{\Lambda}\!\!\!\! d s {\mathcal{L}}(\partial_s \Phi(s),\Phi(s))   \rightarrow \int^{T^\mu} \!\!\!\!\!\! d^4 x {\mathcal{L}}(\partial_\mu \Phi(x),\Phi(x)) 
\label{action:compact}
\end{equation}
This formalism is manifestly covariant (the action is a scalar) as the PBCs (or anti-PBCs) minimize the action at the boundary \cite{Dolce:2009ce,Dolce:tune}. A global Lorentz transformation $x^\mu \rightarrow x'^\mu = \Lambda^{\;\mu}_{\nu} x^\nu $  in the free action (\ref{action:compact})  implies a transformation of the boundary $T^\mu \rightarrow {T'}^\mu = \Lambda^{\;\mu}_\nu T^\nu$. The resulting recurrence  ${T'}^\mu $ actually describes the  four-momentum $\bar k_\mu \rightarrow  {{\bar k}'}_\mu = \Lambda_{\;\mu}^\nu \bar k_\nu$ of the free particle in the new frame,  according to  $\bar k'_{\mu}  T'^{\mu} = 2 \pi$. The geometric quantity $T^{\mu}$ is therefore a contravariant tangent four-vector \cite{Kenyon:1990fx} satisfying the reciprocal of the relativistic dispersion relation: $\bar m^{2}= \bar k_{\mu} \bar k^{\mu} \Leftrightarrow \frac{1}{\Lambda^{2}} = \frac{1}{T^{\mu}} \frac{1}{T_{\mu}}$. Thus, as long as modulations of periodicities are considered, the resulting description of elementary spacetime cycles are fully consistent with special relativity.

In this section we only consider the non-quantum limit of a bosonic massive particle. This corresponds to the fundamental solutions of (\ref{action:compact}), denoted by the bar sign,  $\bar \Phi(s) \propto \exp[-i \bar m s] \rightarrow \bar \Phi(x) \propto \exp[-i \bar k_{\mu} x^{\mu}]$.  Indeed, the boundary at $\Lambda$ and $T^{\mu}$ select the kinematical state of the particle, \ie the BCs select a single Klein Gordon (KG) mode describing the free  boson at rest or in a generic frame, respectively. Thus, in a generic frame,  we see   that in (\ref{action:compact}) such a free fundamental solution $\bar \Phi(x)$ of persistent periodicity $T^{\mu}$ is described by the ordinary KG lagrangian ${\mathcal{L}} \Rightarrow \bar{\mathcal{L}}_{KG} = \frac{1}{2} [\partial_\mu \bar \Phi\partial^\mu\bar\Phi-\bar{m}^{2}\bar{\Phi}^{2}]$.   

Generic interactions are introduced by considering the corresponding periodicity modulations, as prescribed by undulatory mechanics. 
That is, in every point $x=X$, a generic interaction scheme (\eg a 
potential) can be described by the corresponding relativistic variations of four-momentum with respect to the free case $\bar{k}_{\mu}\rightarrow\bar{k}'_{\mu}(X)=e_{\;\mu}^{a}(x)|_{x=X}\bar{k}_{a}$. This implies relativistic modulations of instantaneous four-periodicity $T^{\mu}\rightarrow T'^{\mu}(X)\sim e_{a}^{\;\mu}(x)|_{x=X}T^{a}$. Under the assumption of IP, \emph{every free particle is an inertial reference clock}. Generic interactions are therefore modulations of these elementary clocks and in turns can be encoded in corresponding deformations of the spacetime metric \cite{Birrell:1982ix}, similarly to General Relativity (GR).  In our formalism, modulations of periodicity are in fact realized through PBCs by local deformations of  compact spacetime dimensions. Indeed, in the free action (\ref{action:compact}), the local transformations of reference frame $x^{\mu}\rightarrow x'^{\mu}(x)=x^{a}\Lambda_{a}^{\;\mu}(x)$, where $e^{\;\mu}_{a}(x) = [\partial x'^{\mu}(x)/\partial x^{a}]$ (for simplicity's sake we work in the linear approximation, neglecting self-interactions)  leads to the transformed action with locally deformed metric $\eta_{\mu\nu}\rightarrow g_{\mu\nu}(X)=[e_{\;\mu}^{a}(x)e_{\nu}^{\; b}(x)]_{x=X}\eta_{ab}$ and local boundary $T^{\mu} \rightarrow \Sigma'^{\mu}(X) =\Lambda^{\;\mu}_{a}(x)|_{x=X}T^{a}$:
\begin{equation}
\mathcal{S}^{\Lambda}\simeq\int^{\Sigma'^{\mu}(X)}d^{4}x\sqrt{-g(x)}\mathcal{L}(e_{a}^{\;\mu}(x)\partial_{\mu}\Phi'(x),\Phi'(x))\,.
\label{eq:defom:action:generic:int}
\end{equation}
The resulting fundamental solution  has the typical form of a modulated wave $\bar\Phi'(x)\propto\exp[-i\int^{x_{\mu}}dy^{\mu}\bar {k}'_{\mu}(y)]$ of instantaneous periodicity $T'^{\mu}(x)$. Indeed, it actually describes the local four-momentum $\bar{k}'_{\mu}(x)$ of the interacting bosonic particle. 

It is easy to check that this description of interaction matches the ordinary geometrodynamical description of gravitational interaction of GR. A weak Newtonian interaction in fact corresponds  to the local variations of energy  $\bar{\omega}\rightarrow\bar{\omega}(x)'\sim\left(1+{GM_{\odot}}/{|\mathbf{x}|}\right)\bar{\omega}$, implying through the Planck constant the corresponding modulations of time periodicity $T_{t}\rightarrow T_{t}'(x)\sim\left(1-{GM_{\odot}}/{|\mathbf{x}|}\right)T_{t}$. That is, elementary cycles applied to Newtonian interaction immediately yields two fundamental aspects of GR: redshift and time dilatation, \eg see \cite{Ohanian:1995uu}. By also considering the variations of momentum and the corresponding modulations of wavelength we find that, in (\ref{eq:defom:action:generic:int}), the Newtonian interaction is actually encoded by the linearized Schwarzschild metric. As known, ordinary Einstein's equation follows from this linearized description by considering self-interaction  \cite{Ohanian:1995uu}. 

Note that relativity, if rigorously interpreted, only fixes the differential structure of spacetime (metric), whereas the only requirement for the BCs comes from the variational principle, \ie the BCs must minimize the action at boundary of a relativistic theory.  Einstein's equation is defined modulo boundary terms: we do not know ``what [and where] is the boundary of GR''  \cite{springerlink:10.1007/BF01889475}. We also note that our approach fulfills the holographic principle \cite{'tHooft:1993gx}: the boundary $\Sigma'^{\mu}(x)$ of (\ref{eq:defom:action:generic:int}) is directly related to the local spacetime periodicity $T'^{\mu}(x)$ and therefore, through $\hbar$, it explicitly encodes the kinematics $\bar{k}'_{\mu}(x)$ of the interaction.  
  
 The assumption that every particle is a reference clock yields a geometrodynamical description of gauge interaction, similar to GR as in Weyl's original proposal. Intuitively we want to describe the motion of a particle interacting electromagnetically in terms of the corresponding local transformations of flat reference frame: $dx^{\mu}(x)\rightarrow dx'^{\mu}\sim dx^{\mu} - e dx^{a} \zeta^{\;\mu}_{a}(x)$, ($g_{\mu\nu}(x) \propto \eta_{\mu\nu}$). This can be parametrized by the vectorial field $\bar{A}_{\mu}(x) \equiv \zeta_{\;\mu}^{a}(x)\bar{k}_{a}$ so that, according to our geometrodynamical description, the resulting interaction scheme is the minimal substitution  $\bar{k}'_{\mu}(x)  \sim  \bar{k}_{\mu}- e\bar{A}_{\mu}(x)$ \cite{Dolce:tune}. The  locally rotated boundary of  (\ref{eq:defom:action:generic:int}) leads to modulated solutions of the type $\bar \Phi'(x) \propto \exp[{{-i \bar k_{ \mu} x^{\mu}} + i e \int^{x^{\mu}} d y^{\mu} \bar A_{ \mu} (y) }]$.  The gauge connection (Wilson line) $\bar U(x) = \exp[{ i e \int^{x^{\mu}} d y^{\mu} \bar A_{ \mu}(y) }]$ (postulated in ordinary gauge theory) here is \emph{derived} from local transformations of flat reference frame. Indeed, through the gauge transformed terms $ \bar \Phi(x)=\bar U^{-1}(x) \bar \Phi'(x)$ and $\partial_{\mu} \bar \Phi(x) =\bar U^{-1}(x)D_{\mu} \bar \Phi'(x) $, where $D_{\mu}=\partial_{\mu}-ie\bar{A}_{\mu}(x)$, the modulated periodicity $T'^{\mu}(x)$ of the interacting particle $ \bar \Phi'(x)$ is ``tuned'' to the persistent periodicity $T^{\mu}(x)$ of the free case  $\bar \Phi(x)$. By substituting  $\bar \Phi(x)$  in ${\bar\mathcal{L}_{KG}}$, we see that the modulated solution  $\bar \Phi'(x)$ of the interacting action with locally deformed boundary (\ref{eq:defom:action:generic:int}) is equivalently obtained from the ``tuned'' fundamental lagrangian ${\mathcal{L}} \Rightarrow {\bar\mathcal{L}_{tuned}} = \frac{1}{2} [D_\mu \bar \Phi' D^\mu \bar \Phi' - \bar m^{2} \bar {\Phi'}^{2}]$ in the action (\ref{action:compact}) with persistent boundary $T^{\mu}$. In general \cite{Dolce:tune}, gauge invariant terms are ``tunable'' to fulfill the variational principle at a common boundary $T^{\mu}$. In this way we also justify the kinematic term $-\frac{1}{4}\bar F_{\mu\nu} \bar F^{\mu\nu}$ for $\bar A_{\mu}$, formally obtaining a Yang-Mills theory  (this constrains $\zeta^{\;\mu}_{a}(x)$ to be ``polarized''). We have inferred that local transformations of flat reference frame, inducing local rotations of the boundary, correspond to  transformations of the fundamental solution of (\ref{eq:defom:action:generic:int}) formally equivalent to the internal transformations of ordinary gauge theory (in a sort of parametrization invariance applied to the boundary).

Though relativity fixes the differential structure of spacetime, it allows us to ``play'' with spacetime boundaries as long as the variational principle is fulfilled. Nevertheless, BCs (playing a central role in ``old'' QM) have a  ``marginal'' role in Quantum Field Theory (QFT). For instance, the KG field used for mathematical computations is the more general solution of the KG equation, so that variations of the boundaries and BCs are not considered in the ordinary KG fields. We have introduced a formalism in which the boundary selects the kinematical state of the particle.  Interactions can be therefore described as corresponding local retarded modulations of the spacetime periodicity of a  KG mode from the initial state to the final state, as suggested by undulatory mechanics, instead of creating and annihilating modes with different spacetime periodicities (\ie four-momenta) as in ordinary QFT.

\paragraph*{\bf Correspondence to quantum mechanics}---
By playing with BCs in a consistent way in a relativistic wave theory, it is possible to derive at a classical level the fundamental aspects of relativistic QM  \cite{Dolce:2009ce,Dolce:tune,Dolce:AdSCFT}. The most general bosonic solution  of the free action (\ref{action:compact}) is a vibrating ``cord''\footnote{We use the term ``cord'' to avoid confusion with string theory} with all the harmonics allowed by the PBCs at $T^{\mu}$:  $\Phi(x) = \sum_{n} \phi_n(x) = \sum_{n}  \mathcal N {a^*}_{n}(\bar \mathbf k) \exp[{-i k_{n \mu} x^{\mu}}]$; the KG mode $\bar \Phi$ considered so far is the fundamental harmonic, $\mathcal N$ and $a^{*}_{n}(\bar \mathbf k) $ are the normalization and population coefficients of the harmonics, respectively.  Homogeneous spacetime periodicity $T^{\mu}$ (\ie isolated particle) implies a harmonic quantization of the conjugate energy-momentum spectrum $k_{n \mu} T^{\mu}  = n \bar k_{\mu} T^{\mu} = 2 \pi n$;  that is, time periodicity $T$ implies the quantized harmonic energy spectrum $\omega_{n} = n \bar \omega = 2 \pi n / T$. Since $T^{\mu}$ transforms as a contravariant four-vector, $T$ varies with $\bar \mathbf k$ (relativistic Doppler effect). In a generic reference frame, the resulting energy spectrum of our harmonic system  is therefore $\omega_{n} (\bar \mathbf k) = 2 \pi n  / T (\bar \mathbf k) = n \sqrt{\bar \mathbf k^{2}  + \bar m^{2}}$. This is the same energy spectrum prescribed for free bosons by ordinary second quantization in QFT (after normal ordering\footnote
{Another equivalently allowed possibility is to assume anti-PBCs instead of PBCs, obtaining $\omega_{n}(\bar \mathbf k)  =   (n + \frac{1}{2})  \frac{2 \pi}{ T(\bar \mathbf k)} =   (n + \frac{1}{2})  \bar \omega (\bar \mathbf k) $.}). 

The modulated IP $T'^{\mu}(X)$ of an interacting particle $ \exp[- i\int^{\Sigma'^{\mu}(X)} d x^{\mu} \bar {k}'_{n \mu}(x) ]\equiv \exp[-i2\pi n]$ implies the Bohr-Sommerfeld quantization  $\oint_{X} d x^{\mu} \bar {k}'_{n \mu}(x) = 2 \pi n$, from which it is possible to solve semiclassically fundamental quantum systems \cite{refId0}. This modulated harmonic system is the typical classical system representable locally in a Hilbert space. The modulated harmonics of our vibrating ``cord'' form locally a complete set with inner product $\left\langle \phi'_{n}(X,t_{f})|\phi'_{n'}(X,t_{i})\right\rangle \equiv\int_{V({\mathbf{X})}}\frac{{d\mathbf{x}}}{V(\mathbf{x})}
{\phi'_{n'}}^{*}(\mathbf{x},t_{f})\phi'_{n}(\mathbf{x},t_{i})$ and Hilbert eigenstates $\left\langle X | \phi'_{n}\right\rangle \equiv \phi'_{n}(x)$ ($V(\mathbf{x})$ corresponds to a sufficiently large or infinite number of spatial periods to contain the interaction region, so that $V(\mathbf{x})\simeq V$). Thus, $\bar \Phi'(x)$ is represented by a corresponding Hilbert state $ \left| \Phi' \right\rangle = \sum a_{n} \left| \phi'_{n} \right\rangle$. The composition of more harmonic systems is represented by the tensor product of the Hilbert spaces with a quantum number for every fundamental periodicity; $\{n\}$ corresponds to the $\mathbb S^{1}$ spacetime recurrence\footnote{\textit{E.g.}  spatial spherical symmetry $\mathbb S^{2}$ implies ordinary angular momentum quantization  with two additional quantum numbers $\{l,m\}$), with related harmonic expansion and implicit commutation relations.}. 

The energy-momentum spectrum of the locally modulate harmonic system defines the non-homogeneous Hamiltonian  and momentum operators $\mathcal H'$ and $\mathcal P'_{i}$ as $\mathcal P'_{\mu}\left |\phi'_{n} \right\rangle \equiv k'_{n \mu}  \left|\phi'_{n} \right\rangle$, where $\mathcal{P}'_\mu = \{\mathcal H', - \mathcal{P}'_{i}\}$.  The ``square root'' of the modulated wave equation leads to the spacetime evolution of every harmonic $i \partial_{\mu} \phi'_{n}(x) = k'_{n}(x) \phi'_{n}(x)$ with infinitesimal unitary evolution operator $\mathcal U(x+d x) = \exp[-i \mathcal P'_{\mu}(x) dx^{\mu}]$. Thus, the resulting time evolution  of our modulated vibrating ``cord'' in Hilbert notation is described by the ordinary Schr\"odinger equation $i  \partial_{t} \left|\Phi' \right\rangle = \mathcal H' \left|\Phi' \right\rangle$. Moreover, the ordinary commutation relations of QM are directly \emph{derived} from  the constraint of IP. In analogy with Feynman's demonstration \cite{Feynman:1942us}, the expectation value of a total derivative $\partial_{i} \mathcal F(x)$ for generic Hilbert states, after integration by parts and considering that the boundary terms cancel owing IP, yields the same commutators of ordinary QM: $ [\mathcal F(x),\mathcal{P}'_{i}] = i \partial_{i} \mathcal F(x) $, and $ [x_{j},\mathcal{P}'_{i}] = i \delta_{i j} $ for $\mathcal F(x)=x_{j}$  \cite{Dolce:2009ce,Dolce:tune}.

 This correspondence to ordinary relativistic QM is also proved by the fact that the \emph{classical} evolution of our modulated harmonic system is mathematically described by the ordinary Feynman path integral  \cite{Dolce:2009ce,Dolce:tune}
 \begin{equation}
\mathcal{Z}=\int_{V} {\mathcal{D}\mathbf{x}} e^{i \mathcal{S}'(t_{f},t_{i})}\,.
\label{eq:Feynman:Path:Integral}
\end{equation}
The resulting lagrangian $\mathcal{L}' = {\mathcal P}'_{i}  x_{i}- \mathcal H'$ associated to $\mathcal S'$ is, by construction, the classical action of the corresponding interaction scheme, according to the ordinary formulation. This remarkable result has an intuitive, purely classical interpretation. Without relaxing the classical variational principle, in our cylindric geometry of topology $\mathbb S^{1}$ the classical evolution of $\Phi'(x)$ is the result of the interference of all the (potentially infinite) possible classical paths with different windings numbers between its two extremal configurations. By extending all the volume integrals to an infinite number of periods ($V\rightarrow \infty$) we obtain formally the same equations of relativistic QM, and thus of our modern description of nature. It is straightforward to check that our formalism provides a correct classical limit for $\hbar \rightarrow 0$. The harmonics of $\Phi'(x)$ are interpreted as quantum excitations of the system.  The corpuscular limit corresponds to $T \rightarrow 0$, so that only the fundamental energy level (KG mode) is populated (infinite energy gap).   IP applied to fermions represents a natural realization of the \emph{Zitterbewegung} model, providing a semi-classical description of the spin and the Dirac equation \cite{Hestenes:zbw:1990,Trzetrzelewski:2013xia}, as well as interesting reconsiderations of supersymmetry foundational aspects \cite{CasalbuoniZitterSusy}. 

The modulated classical evolution (\ref{eq:Feynman:Path:Integral}) in the case of gauge  interaction turns out to be described by the ordinary Feynman path integral of scalar QED (similarly, the scattering matrix describes the modulations of the harmonic modes in Hilbert notation) \cite{Dolce:tune}. 
 Such a semiclassical description  is confirmed by recent progresses in integrable theories with fundamental analogy to our theory, such as light-front-quantization \cite{Zhao:2011ct} (where PBCs are the quantization condition), Twistor theory \cite{ArkaniHamed:2009si} (``cyclic coordinates'' attached to every spacetime point) and AdS/CFT (discussed below).  
 The modulated PBCs applied locally to the modulo of the gauge transformation (Wilson loop with generic winding number) $ \exp[ i e \int^{\Sigma'^{\mu}(X)} d x^{\mu} A_{ \mu}(x) ] = \exp[ i \pi n]$ implies the Dirac quantization for magnetic monopoles $ \oint_{X} d x^{\mu} e A_{ \mu}(x)   =   \pi n$. 
For example, as described in a forthcoming paper, IP in gauge transformations 
$A_{ \mu}(x) \rightarrow A_{ \mu}(x) - e \partial_{\mu} \Theta(x)$ implies that the corresponding Goldstone mode $\Theta(x)$, being a phase of a ``periodic phenomenon'', is periodic and can only vary by discrete amounts: $ e \Theta(x^{\mu}) =  e \Theta(x^{\mu} + T^{\mu}) + \pi n \Rightarrow \delta \Theta(x) = n /2 e$. Thus, through Stokes' theorem, this means that the magnetic flux along an orbit $\Sigma$ characterized by pure gauge is quantized and the current cannot decay smoothly: $\int_{\mathcal{S}_{\Sigma}}\mathbf{B}(\mathbf{x},t)\cdot d\mathbf{S} 
= \oint_{\Sigma_{\mathcal{}}}\mathbf{A}(\mathbf{x},t)\cdot d\mathbf{x}
  =  \oint_{\Sigma}\mathbf{\nabla}{{\theta}}(\mathbf{x})\cdot d\mathbf{x}= n / 2 e$.  
 We have therefore obtained the description of superconductivity given in \cite{Weinberg:1996kr}. Superconductivity has historically originated the Higgs mechanism, in our description the corresponding gauge symmetry breaking mechanism can be interpreted as a quantum-geometrodynamical effect, see \cite{Dolce:tune,Dolce:SuperC}. 

We propose the following interpretation of the mathematic results above. This formal correspondence suggests a possible statistical origin of QM associated to the fact that, typically, the Compton periodicity of elementary particles is extremely fast (except neutrinos) with respect to the present experimental time resolution $\sim 10^{-17} s$. For instance, the recurrence of a simple electrodynamical system is fixed by the electron Compton time: $\Lambda_{e}= 8.09329972 \times 10^{-21} s$. As also noted by 't Hooft \cite{'tHooft:2001ar}, ``there is a deep relationship between a particle moving [very fast] along a circle and the quantum harmonic oscillator'' (\ie the basic element of QFT). Indeed, such a relativistic particle moving very fast along a circle can only be described statistically (Hilbert notation) as a fluid (continuity equation) of total density one 
 (Born rule)\footnote{It is interesting to note that this is analogous to the statistical description of the outcomes of a die rolling very fast. In a die the outcomes can be predicted by observing the motion with a sufficient resolution in time (an imaginary observer with infinite time resolution has no fun playing dice, ``God does not play dice'', A. Einstein.) }.  
Since in our statistical (Hilbert) description  of a ``periodic phenomenon'' (reference clock) the phase cannot be observed, to determine the energy and define time with good accuracy  we must count a large number of ``ticks'', according to $|\exp[-i{\bar \omega t}]|= \exp[-i({\bar \omega t}+ \pi)] = \exp[-i(\bar \omega + \Delta\bar{\omega})  t ] = \exp[-i \bar \omega  ( t + \Delta t)]$. This corresponds to Heisenberg's relation $\Delta\bar{\omega}\Delta t\ge  1/2$ \cite{Dolce:2009ce,Dolce:tune}.
 Note that we have not introduced any local-hidden-variable in the theory. The assumption of IP is the only quantization condition and represents an element of non-locality (namely ``complete coherence''\footnote{The quantum recurrence is destroyed by the dissipative thermal noise (chaotic interactions) \cite{2013arXiv1304.6295F}. This correctly describe the quantum to classical transition.  Cyclic dynamics suggests a physical interpretation of the ``mathematical trick'' of Euclidean time periodicity imposed to quantize statistical systems and of Wick's rotation \cite{Dolce:SuperC}: the assumption in IP is in general a quantization condition for a system at the equilibrium.}, consistent with relativistic causality). The formal correspondence to ordinary QM suggests that the theory could violate Bell's inequalities as ordinary QM, and thus a possible deterministic nature at time scales smaller than the Compton time of the system.

\paragraph*{\bf Applications to modern physics}--- 
Elementary spacetime cycles represent a new interpretational key of fundamental aspects of modern physics, yielding elements of the mathematical beauties of extra-dimensional or string theories. The Compton periodicity of the worldline parameter enters into the equations of motion of the free field $\Phi(x)$ in perfect analogy with the cyclic eXtra-Dimension (XD) of a Kaluza-Klein (KK) theory, \cite{Klein:1926tv,Kaluza:1921tu}. If we have a KK massless theory $dS^{2} = dx_{\mu} d x^{\mu} - d s^{2} \equiv 0$ with cyclic XD $s$  and compactification length $\Lambda$,  and we identify the cyclic XD with the Minkowskian worldline parameter, $d s^{2} = dx_{\mu} d x^{\mu} $, we obtain exactly our purely 4D theory with Compton worldline periodicity $s \in (0,\Lambda]$ and thus, through Lorentz transformations, elementary spacetime cycles. We address this identification by saying that the XD is \emph{Virtual} (VXD) \cite{Dolce:AdSCFT}. For instance, the rest energy spectrum of our harmonic system, or equivalently of a second quantized KG field,  is the analogous of the KK mass tower $m_{n} = n \bar m = 2 \pi n / \Lambda \equiv \omega_{n} (0) = n\bar \omega (0) = 2 \pi n / T(0)$\footnote{This rest spectrum is observed in quantum phenomena such as superconductivity and nanotubes \cite{deWoul:2012ed}.}.  In this case the KK modes are \emph{virtual}: they are not independent KK particles of different masses, they describes the energy excitations of the same 4D harmonic system, \ie the quantum excitations of the same system. The holographic approach in XD theories actually corresponds to the same collective\footnote{In the holographic propagator $\Pi^{Holo}(\bar k^{2})$ the effective propagation of the all KK modes is in fact described in a collective way in terms of the same fundamental $\bar k_{\mu}$ fixed by the source field.} description of the KK modes, but in addition to this the heavy modes are integrated out in terms of the source field $\phi_{\Sigma}(x) \sim \bar \Phi(x)$.  The result of the holographic approach is thus equivalent to an effective description of the VXD: $ \mathcal{S}'^{Holo}_{\Phi|_{\Sigma}=e\phi_{\Sigma}}(s_{f},s_{i}) \sim {\mathcal{S}}^{VXD}(s_{f},s_{i})+\mathcal{O}(E^{eff}/\bar{m})$ (with implicit source term) \cite{Dolce:AdSCFT,Casalbuoni:2007xn}. We note that, similarly to light-front-quantization \cite{Honkanen:2010nt}, Hilbert notation with Schr\"odinger evolution $i  \partial_{s} \left|\Phi'\right\rangle = \mathcal M \left|\Phi' \right\rangle$  and implicit commutation relation $[s,\mathcal{M}] = i $ between mass operator $\mathcal M \left |\phi_{n} \right\rangle \equiv m_n  \left|\phi_{n} \right\rangle$ and XD $s$ can also be applied to KK theories \cite{Dolce:AdSCFT}.

 An example \cite{Dolce:AdSCFT,Dolce:graphene} of this geometric interpretation of the particle's mass as Compton proper-time periodicity, as well as of the analogy between XD geometry and quantum behavior, comes from graphene physics. In graphene the electrons behave as 2D massless particles. However, as a dimension is curled up to form a nanotube, the electrons at rest with respect to the axial direction (rest frame) have  residual cyclic motions along the radial direction: the residual proper-time periodicity of  corresponds to the Compton time defining the effective mass of the electrons in nanotubes, \cite{deWoul:2012ed,RevModPhys.79.677,Dolce:AdSCFT,Dolce:graphene}. A similar example is given by graphene bilayer, in which the effective mass  is originated by the residual proper-time periodicity associated to the cyclic motion of the electrons between the two layers at different potentials \cite{2011arXiv1103.1663Z}. 

By means of the duality to XD theories, the generic interaction scheme $\bar{k}'_{\mu}(x)$ is equivalently encoded in a corresponding deformed\footnote{In the case of finite Compton periodicity, dilatons or softwalls should be included as well as mixing terms $\mathcal O(dx^{\mu} ds)$.} VXD metric $d S^{2} \simeq g_{\mu\nu} d x^{\mu} d x^{\nu} - d s^{2}$. As confirmation of our description of gauge interaction, it can be shown that the corresponding spacetime geometrodynamics discussed above are equivalently encoded in a \emph{virtual} Kaluza-like metric (Kaluza's miracle) \cite{Kaluza:1921tu}. 
By combining the correspondence with relativistic QM \cite{Dolce:2009ce}, the geometrodynamical formulation of interactions \cite{Dolce:tune} and the dualism to XD theories, we obtain that elementary cycles pinpoints the  fundamental classical to quantum correspondence of Maldacena's conjecture and holography \cite{Witten:1998zw,Gherghetta:2010cj}.  This \emph{implies} that the holographic representation of the  classical configurations of $\Phi'$ in a curved XD background describes the low energy quantum behavior of the corresponding interaction scheme \cite{Dolce:AdSCFT}
\begin{equation}
\int_{V} {\mathcal{D}\mathbf{x}} e^{ i \mathcal{S}'(t _{f},t_{i})} \leftrightsquigarrow  e^{ i \mathcal{S}'^{Holo}_{\Phi|_{\Sigma}=e\phi_{\Sigma}}(s_{f},s_{i})}\,.\label{VXD:QFT:corr}
\end{equation} 
  
Another example of this classical to quantum correspondence is given by the Quark-Gluon-Plasma (QGP) freeze-out. We assume that the QGP is described by the classical Bjorken hydrodynamical model \cite{Magas:2003yp}. 
This means that during the freeze-out the energy decays exponentially with the proper-time $s$. Thus, neglecting masses  ($\omega \sim k$),  the four-momentum has  exponential conformal decay $\bar{k}_{\mu}\rightarrow\bar{k}'_{\mu}(s)\simeq e^{-K s}\bar{k}_{\mu}$\footnote{In QCD  thermodynamics \cite{Satz:2008kb},  $K$ is the cooling gradient (Newton's law).}. 
This equivalently means that the spacetime periodicity has  exponential conformal
dilatation $T^{\mu}\rightarrow T'^{\mu}(s)\simeq e^{K s }T^{\mu}$.
According to our geometrodynamical description, this interaction scheme is encoded in the warped metric $ds^{2}=e^{-2K s}dx_{\mu}dx^{\mu}$, resulting from the deformation $dx_{\mu}\rightarrow dx'_{\mu}(s)\simeq e^{-ks}dx_{\mu}$, and in turns in the
\emph{virtual} AdS metric $dS^{2}\simeq e^{-2 K s }dx_{\mu}dx^{\mu}-ds^{2}\equiv0$. 
The time periodicity of the fields during the QGP freeze-out turns out to be the conformal parameter $T(s) = e^{K s }/K = 2 \pi / E(s)$.  In agreement with the AdS/CFT dictionary, this naturally describes, by means of the Planck constant, the inverse of the energy during the freeze-out. Infinite VXD means that the system has infinite Compton recurrence. Indeed, according to our classical to quantum correspondence, the modulated classical configurations of a harmonic system in such a unbounded warped metric encode the quantum behavior of a massless quantum system, \ie of a conformal theory. In particular the warped modulations of the harmonic solution leads to asymptotic freedom, as shown in \cite{Pomarol:2000hp,ArkaniHamed:2000ds}.   

 A massive system has finite Compton periodicity, \ie compact VXD. Indeed, this breaks the conformal invariance. It is known from AdS/QCD that the collective modulated harmonics in this compact warped configuration qualitatively matches the quantum behaviors and mass spectrum of the hadrons (improvements of this model are mentioned in \cite{Dolce:AdSCFT}). Similarly to Veneziano's original idea, in our description the hadrons are indeed collective energy (quantum) excitations, \ie \emph{virtual} KK modes, of the same fundamental vibrating ``string'' \cite{FoundString}. This allows us to introduce another remarkable property of the theory. From the free rest action in (\ref{action:compact}) we see that the Compton wordline recurrence of elementary particles allow us the possibility to define of a novel, minimal string theory based on a single compact world-parameter and a purely 4D target spacetime. From a mathematical point of view, the compact worldline parameter of the theory plays a role similar to both the compact worldsheet parameter of ordinary string theory and of the XD  of the KK theory. For this reason the theory inherits fundamental aspects of both string and XD theories, avoiding the introduction of any unobserved XD. As we will show in forthcoming papers, this yields interesting analogies to Veneziano's amplitude \cite{Afonin:2011hk},  Virasoro's algebra \cite{Dolce:tune}, phenomenological properties of Randall-Sundrum models and implications to quantum gravity. The compact 4D target spacetime has also important justifications in QCD confinement \cite{PhysRevD.9.3471}.

\paragraph*{\bf Conclusions}---
Pure quantum systems are characterized by IP, so that isolated elementary particles can be regarded as reference clocks ``linking time to the particle's mass'' \cite{Lan01022013}. We infer that a formulation of physics in terms of intrinsically cyclic spacetime dimensions is fully consistent with the theory of relativity \cite{Dolce:2009ce,Dolce:tune,Dolce:AdSCFT}. In fact, relativity fixes the spacetime differential structure (metric) leaving the freedom to play with BCs, as long as the variational principle is fulfilled. This allows us to encode undulatory mechanics directly into the spacetime geometry, enforcing the wave-particle duality and the local nature of relativistic spacetime. We have shown that ordinary quantum behaviors are consistently derived from corresponding classical cyclic dynamics. This  provides an intriguing unified description of fundamental aspects of modern physics, including a geometrodynamical description of gauge interaction analogous to gravity, and pinpoints the fundamental classical to quantum correspondence characteristic of important integrable theories.

 \paragraph*{Acknowledgments:} 
 Parts of this work have been presented in recent international conferences \cite{Dolce:cycle,Dolce:ICHEP2012,Dolce:Dice2012,Dolce:TM2012}.

\bibliographystyle{apsrev}
\bibliography{../cycles}

\end{document}